\documentclass[twocolumn,prb,showpacs]{revtex4}
\usepackage{graphicx}
\usepackage{dcolumn}
\usepackage{bm}
\usepackage{amsmath,amssymb}

\begin{document}

\preprint{}
\title{Majorana current induced by charge current in Kitaev magnet/graphene bilayers}
\author{Takehito Yokoyama}
\affiliation{Department of Physics, Institute of Science Tokyo, Tokyo 152-8551,
Japan
}
\date{\today}

\begin{abstract}
We investigate Majorana current induced by charge current in Kitaev magnets coupled with graphene through the mechanism by which Majorana fermions in the Kitaev magnet are dragged by electrons in graphene.
We calculate Majorana current density based on the perturbation with respect to a Kondo type Majorana-electron coupling and obtain an analytical expression of the Majorana current induced by charge current, indicative of electronic control of charge-neutral Majorana fermions. We also discuss Majorana heat current induced by charge current and give estimations of their magnitudes.

\end{abstract}

%\pacs{}
\maketitle

%\section{Introduction}

Quantum spin liquids (QSLs) constitute a class of strongly correlated phases characterized by the absence of conventional symmetry-breaking order and the presence of long-range entanglement and fractionalized excitations.\cite{XGWen2004,Zhou2017QSL,SavaryBalents2017,Wen2019ExpQSL,KnolleMoessner2019,Hermanns2018,Broholm2020,Takagi2019,Trebst2022,Nasu2024,Matsuda2025} A paradigmatic example is the exactly solvable honeycomb model introduced by Kitaev, in which spin-1/2 moments interact via bond-dependent Ising exchange interactions.\cite{Kitaev2006} The model realizes a QSL with emergent Majorana fermions coupled to a static  $\mathbb{Z}_2$ gauge field, providing a rare setting where fractionalization and topological order can be treated analytically. When time-reversal symmetry is broken, the system enters a chiral phase supporting non-Abelian Ising anyons, thereby establishing a direct connection between quantum magnetism and topological quantum computation, where quantum information is encoded and manipulated through the braiding of such excitations.\cite{Kitaev2003,Freedman2003,DasSarmaFreedmanNayak2006,Nayak2008,SternLindner2013,LahtinenPachos2017,FieldSimula2018}

Motivated by this framework, considerable effort has been devoted to identifying material realizations of dominant Kitaev interactions. In spin-orbit-entangled Mott insulators with edge-sharing octahedra, such as $\alpha$-RuCl$_3$ and the honeycomb iridates A$_2$IrO$_3$(A = Na, Li), strong spin-orbit coupling and crystal-field effects generate effective $J=1/2$ moments with highly anisotropic exchange interactions. \cite{Kim2015RuCl3,Yadav2016RuCl3,Jackeli2009,Chaloupka2010} These Kitaev magnets generally host additional Heisenberg and off-diagonal interactions, which stabilize magnetically ordered ground states at low temperatures. Nevertheless, a range of experimental observations  -including broad continua in inelastic neutron scattering and unconventional thermal transport-  indicates proximity to Kitaev QSL physics.\cite{Nasu2017,VinklerAviv2018,KasaharaNature2018,KasaharaPRL2018,Ye2020}
 In particular, external perturbations such as magnetic fields can suppress magnetic order and induce disordered phases with possible topological character. Notably, a half-integer quantized thermal Hall response reported in $\alpha$-RuCl$_3$ has been interpreted as evidence for a chiral Majorana edge mode, consistent with the non-Abelian phase anticipated from the Kitaev model.\cite{KasaharaNature2018,KasaharaPRL2018}

While bulk Kitaev magnets provide a platform for realizing fractionalized excitations, a central challenge for topological quantum computation lies in their detection and manipulation. 
Majorana fermions in Kitaev QSLs are charge neutral, and their dynamics are directly induced by magnetic field or temperature gradient.\cite{Nasu2017,YaoLee2011,Minakawa2020,Kato2025}
On the other hand, electronic control of Majorana fermions is challenging. Since electronic control of Majorana fermions is in principle possible indirectly through charged particles such as electrons, heterostructures that couple Kitaev magnets to itinerant electron systems offer a promising route.\cite{Aasen2020} Interfaces between graphene and $\alpha$-RuCl$_3$ have been proposed to realize a Kitaev-Kondo lattice, in which Dirac electrons hybridize with the fractionalized excitations of the spin liquid.\cite{Carrega2020,Mazzilli2023,Ermakov2026} Such coupling enables neutral Majorana modes to acquire charge-sensitive signatures, thereby opening pathways for electrical detection and control. Moreover, the tunability of graphene via gating, strain, and moire engineering provides external knobs to access regimes that are difficult to achieve in bulk systems
%including possible heavy Fermi liquid behavior and non-Fermi-liquid phases arising from strong hybridization.

More broadly, graphene/Kitaev magnet heterostructures offer a versatile platform to explore the interplay between topology, strong correlations, and itinerant electrons. The combination of proximity-induced interactions and high-mobility carriers raises the possibility of engineering devices in which Majorana excitations emerging from a proximate spin liquid can be manipulated using established mesoscopic techniques. 
%In conjunction with superconducting proximity effects, such systems may provide a route toward implementing braiding operations and other key elements required for TQC within a solid-state setting.

In this paper, we investigate Majorana current induced by charge current in Kitaev magnets coupled with graphene through the mechanism by which Majorana fermions in the Kitaev magnet are dragged by electrons in graphene.
We calculate Majorana current density based on the perturbation with respect to a Kondo type Majorana-electron coupling and obtain an analytical expression of the Majorana current induced by charge current, indicative of electronic control of charge-neutral Majorana fermions. We also discuss Majorana heat current induced by charge current and give estimations of their magnitudes.

Now, let us derive Majorana current induced by an applied electric field in a Kitaev magnet/graphene heterostructure.\cite{Zhou2019RuCl3b,Mashhadi2019b} To this end, we consider the mechanism by which Majorana fermions in the  Kitaev magnet are dragged by electrons in graphene. 
The calculation is similar to that for magnon-drag thermoelectric transport.\cite{Miura2012}
We assume that the localized spins in a Kitaev magnet and the electrons in graphene interact with each other through a Kondo coupling.
The Hamiltonian is composed of four parts: $H = H_K + H_e +H_J +H_A$.
The Kitaev Hamiltonian $H_K$ reads\cite{Kitaev2006,Nasu2024}
\begin{eqnarray}
{H}_K = -J_K \sum_{\langle i,j \rangle^\gamma} s_i^\gamma s_j^\gamma.
\end{eqnarray}
where $s_i^\gamma$ is the spin operator at the site of a honeycomb lattice $i$ with spin component $\gamma=x ,y, z$ and $J_K$ is the Kitaev interaction between nearest neighbors. The sum is taken over nearest neighbors in bond direction $\gamma$ ($=x, y, z$).
We can represent the Kitaev Hamiltonian by means of Majorana fermions.\cite{Kitaev2006,Nasu2024}
See Appendix for details of the derivation of Majorana representation of the Kitaev Hamiltonian.

Then, assuming the ground state, the Kitaev Hamiltonian  near the K point can be reduced to
\begin{align}
H_{K} = \psi^\dagger v (-q_y \rho_x + q_x \rho_y)\psi
\end{align}
with $\psi^\dagger=(c_{\boldsymbol{k}A}^\dagger,c_{\boldsymbol{k}B}^\dagger)$. The annihilation operator at A(B) sublattice is denoted by  $c_{\boldsymbol{k}A(B)}$ and $\rho$ represents the Pauli matrix acting on the sublattice space.
The Kitaev Hamiltonian possesses both time-reversal and inversion symmetries and thus the contributions from the K and K' points are identical. Therefore, we here focus on the contribution from the K point.
The corresponding retarded Majorana Green's function $G^r_{\bm{q}}(\omega)$ reads
\begin{equation}
    G^r_{\bm{q}}(\omega) = (\omega + i\alpha - H_K)^{-1} = \frac{\omega + i\alpha + v(q_y \rho_x - q_x \rho_y)}{(\omega + i\alpha)^2 - v^2 q^2}.
\end{equation}
The broadening parameter $\alpha \sim 0.1-1$ meV\cite{Hirobe2017} stems from non Kitaev interactions, phonon scatterings, impurity scatterings or interaction with electrons.
The velocity operator $v_x$ is given by
\begin{equation}
    v_x = \frac{\partial H}{\partial q_x} = v \rho_y.
\end{equation}

Regarding electrons, let us consider an effective Hamiltonian for graphene near the K point $H_e$,\cite{CastroNeto2009}
\begin{equation}
    H_e = \psi_e^\dagger v_e (k_x \tau_x + k_y \tau_y)\psi_e
\end{equation}
with $\psi_e^\dagger=(a_{\boldsymbol{k}A}^\dagger,a_{\boldsymbol{k}B}^\dagger)$. The annihilation operator at A(B) sublattice is denoted as $a_{\boldsymbol{k}A(B)}$ and $\tau$ denotes the Pauli matrix acting on the sublattice space.
The Hamiltonian for graphene also possesses both time-reversal and inversion symmetries and hence the contributions from the K and K' points are identical. Therefore, we focus on the contribution from the K point.

The retarded electron Green's function $g_{\bm{k}}^r(E)$ is given by
\begin{equation}
    g_{\bm{k}}^r(E) = (E+i\delta - H_e)^{-1} = \frac{E+i\delta - v_e (k_x \tau_x + k_y \tau_y)}{(E+i\delta)^2 - v_e^2 k^2}.
\end{equation}
The broadening parameter $\delta \sim 0.1-10$ meV\cite{Hong2009} comes from electron-electron interaction, electron-phonon interaction, impurity scatterings or interaction with Majorana fermions.
The velocity operator $v_{ex}$ reads
\begin{equation}
    v_{ex} = \frac{\partial H_e}{\partial k_x} = v_e \tau_x.
\end{equation}
The Kondo coupling between local spins and electron spins read
\begin{equation}
H_J = J\sum_{i, \gamma}  \psi_e^\dagger \sigma^\gamma_i \psi_e s^\gamma_i.
\end{equation}
Here, $\sigma$ denotes the Pauli matrix for the electron spin.

%\begin{equation}
%\boldsymbol{s}_i^e = c_{i, \sigma}^\dagger \boldsymbol{\sigma}_{\sigma \sigma'} c_{i, \sigma'} ,
%\end{equation}

We describe  an electric field along $x$-direction by the vector potential $A_x$: $E_x = -\dot{A}_x= i \Omega {A}_x$, which can be included in the Hamiltonian as
\begin{equation}
H_A = \psi_e^\dagger e v_{ex} A_x \psi_e. 
\end{equation}

We investigate Majorana current with respect to the second order of $H_J$. 
Let us show how the perturbation theory works for Majorana fermions in our system, taking $\langle c_i c_j \rangle$ as an illustrative example.
Since Majorana fermions and electrons are decoupled for $J=0$, the expectation value of $c_i c_j$ for the second order of $H_J$ can also be factorized as
\begin{align}
 J^2 \langle c_i c_j \ a_{k \lambda}^\dagger \sigma_{\lambda \lambda_1}^\gamma a_{k \lambda_1} s_k^\gamma a_{l \lambda_2}^\dagger \sigma_{\lambda_2 \lambda_3}^{\gamma'} a_{l \lambda_3} s_l^{\gamma'} \rangle \nonumber \\
 = J^2 \langle c_i c_j c_k c_l \rangle \langle b_k^\gamma b_l^{\gamma'} \rangle \langle a_{k \lambda}^\dagger \sigma_{\lambda \lambda_1}^\gamma a_{k \lambda_1} a_{l \lambda_2}^\dagger \sigma_{\lambda_2 \lambda_3}^{\gamma'} a_{l \lambda_3} \rangle \nonumber \\
    = -i J^2 \langle c_i c_j c_k c_l \rangle \langle a_{k \lambda}^\dagger \sigma_{\lambda \lambda_1}^\gamma a_{k \lambda_1} a_{l \lambda_2}^\dagger \sigma_{\lambda_2 \lambda_3}^\gamma a_{l \lambda_3} \rangle
\end{align}
where we have used  $s_i^\gamma = i b_i^\gamma c_i$. 
Here, the average is taken for the density matrix with $J=0$ and we have focused on diagonal part ofl spin-spin correlations $\langle b_k^\gamma b_l^{\gamma'} \rangle = i \delta_{\gamma \gamma'} u_{kl}=- i \delta_{\gamma \gamma'}$ \cite{Carrega2020,Ermakov2026}, assuming the ground state.
Each average can be calculated by using the standard Green's function method.
Also, we have an additional factor $-i$ in this perturbation. 
Note that since the Hamiltonian for graphene Eq.(5) is spinless and $(\sigma^\gamma)^2=1$, the Pauli matrix $\sigma$ does not appear in  the calculation in second order of $H_J$ below.

%---------------------------------------------------
%	Fig. 1: 
%---------------------------------------------------
\begin{figure}[htb]
%\begin{figure*}[htb]
\begin{center}
\includegraphics[clip,width=8.0cm]{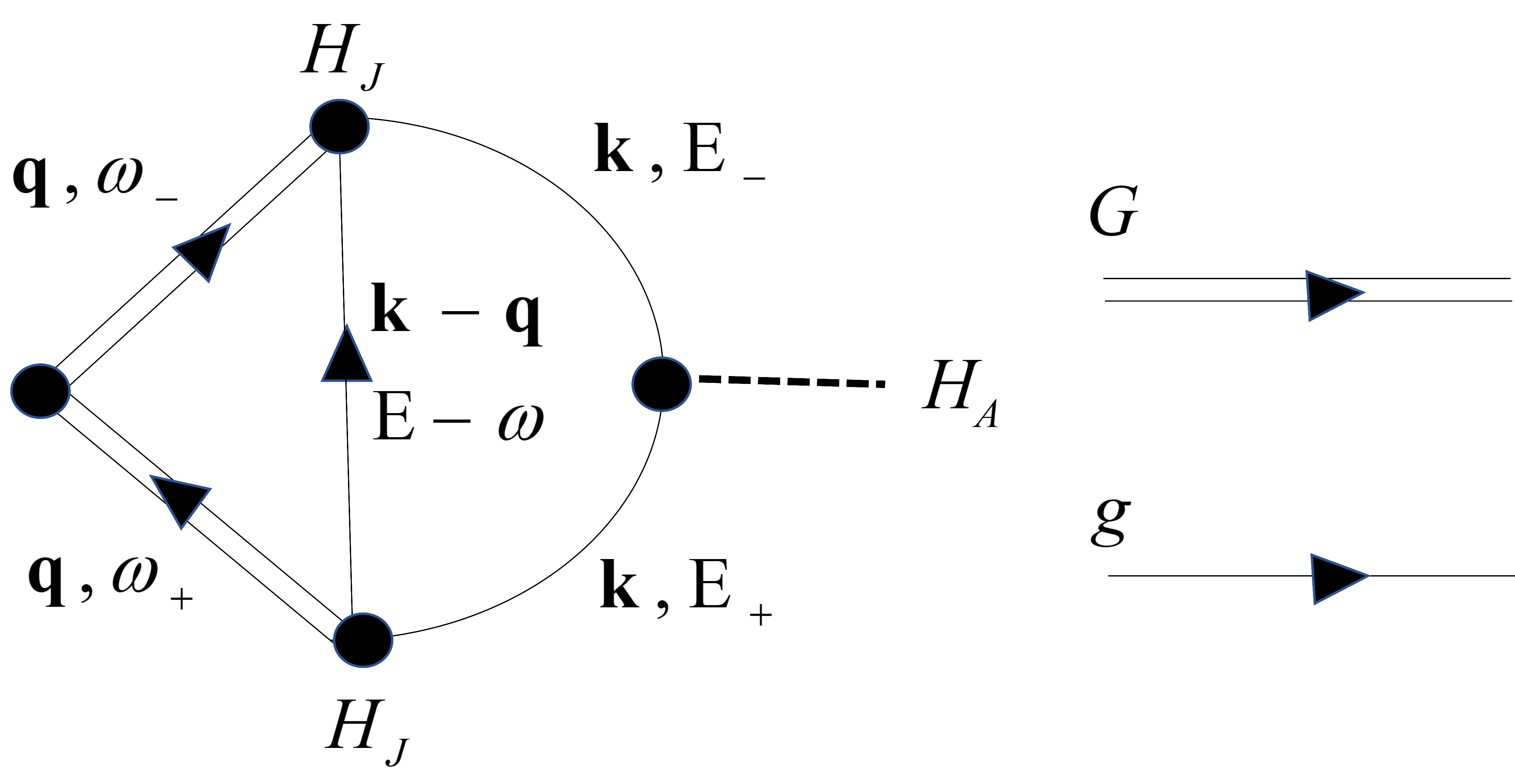}
\end{center}
\caption{
The diagrammatic representations of the Majorana current to first and second orders in ${H_{A}} $ and  ${H_J} $, respectively. Two parallel solid, solid and dotted lines represent Majorana Green's function, electron Green's function and $H_A$, respectively. The left dot between the two parallel solid lines in the left side of the diagram denotes the Majorana current vertex.}
\label{f1}
\end{figure}

The Majorana current density is given by the expectation value of velocity operator which can be calculated by means of the Green's function technique: $j=\langle v_x \rangle = 4 i \text{Tr} \, v_x G^<_{\bm{q}}$.
The trace is taken over momentum (${\bm{q}}$ and ${\bm{k}}$) and sublattice ($\rho$ and $\tau$) degrees of freedom.
We calculate $j$ with respect to the first order of $H_A$ and second order in $H_J$, which is described diagrammatically in Fig. 1, leading to
\begin{equation}
j \cong 4 v \text{Tr} \, \rho_y [G_{\bm{q}} (\omega_-) \Sigma_{\bm{q}}(\omega) G_{\bm{q}}(\omega_+)]^<.
\end{equation}
We here define $\omega_\pm = \omega \pm \Omega/2$ and $E_\pm = E \pm \Omega/2$.
The factor 4 comes from the valley degeneracy of Majorana fermions in the Kitaev magnet and electrons in graphene.

The lesser component can be obtained by using the Langreth theorem:\cite{HaugJauho1997} 
\begin{equation}
    [G_{\bm{q}}(\omega_-) \Sigma_{\bm{q}}(\omega) G_{\bm{q}}(\omega_+)]^< = G^r \Sigma^r G^< + G^r \Sigma^< G^a + G^< \Sigma^a G^a.
\end{equation}

The lesser Majorana Green's function is given by
\begin{equation}
    G_{\bm{q}}^<(\omega) = f(\omega) (G_{\bm{q}}^a(\omega) - G_{\bm{q}}^r(\omega))
\end{equation}
where $f(\omega)$ is the Fermi distribution function.
$G^a$ and $g^a$ are obtained by replacing $\alpha$ by $-\alpha$ in $G^r$ and $\delta$ by $-\delta$ in $g^r$, respectively.
One can calculate the Majorana current density perturbatively by means of these Green's functions.
The details of the calculation are given in Appendix. 

We now assume $\alpha \ll v q \ll T \ll \delta$.
Evaluating the pole at $\omega =  v q + i\alpha$ (due to the Majorana condition $c_{\boldsymbol{k}}^\dagger = c_{-\boldsymbol{k}}$, we only consider the positive-energy branch of the Majorana dispersion\cite{Nasu2015}  and hence the contribution from the pole at Re$\omega>0$),
the Majorana current density $j = \langle v_x \rangle$ is approximated as
\begin{align}
    j &= -4 i v \frac{ v J^2 e \Omega A_x}{2 \pi \alpha \delta} \sum_{\bm{q}} \frac{q_x^2}{(v q)^2} \left. \left( f(\omega) + \frac{d}{d\omega} (\omega n(\omega)) \right)\right|_{\omega=vq} \nonumber \\
    &= \frac{ J^2 e T^2}{24  \alpha \delta v^2} E_x.
\end{align}
This is the central result of this paper.
We find that the Majorana current is induced by an applied electric field in graphene.
The Majorana current density is proportional to $T^2$, which indicates that Majorana fermions are thermally excited.

Let us estimate the magnitude of the Majorana current density.
For $\delta \sim 10 \, \text{meV}$,\cite{Hong2009} $T \sim 1 \, \text{meV}$, 
$\alpha \sim 0.1 \, \text{meV}$,\cite{Hirobe2017}
$J \sim 10 \, \text{meV} \cdot \text{\AA}$,
$v \sim 10^3 \, \text{m/s}$, and
$E_x \sim 10^4 \, \text{V/m}$,
we obtain the magnitude of the Majorana current density as
\begin{align}
    \frac{J^2 e T^2}{24 \alpha \delta v^2 \hbar^3} E_x  \sim 1 \times 10^{18} \, \text{m}^{-1} \cdot \text{s}^{-1}.
\end{align}
Here, $v$ is estimated as $v \sim 3 J_K a/\hbar \sim 10^3 \, \text{m/s}$ for $J_K=1$meV and the lattice constant $a=3$ \AA.
For comparison, the drift velocity for graphene is given as $v_d = \mu E$ with the mobility $\mu$.
For $\mu \sim  1 \, \text{m}^2/\text{V} \cdot \text{s}$ and $E \sim 10^4 \, \text{V/m}$, we have 
$v_d \sim 10^4 \text{m/s}$.\cite{Gosling2021}
For electron density $n \sim  10^{16} \, \text{m}^{-2}$, the particle current density is estimated as 
$n v_d \sim 10^{20} \, \text{m}^{-1} \cdot \text{s}^{-1}$.
The magnitude of Majorana current density is two times smaller than that of electron current density in graphene.

Majorana fermions are charge neutral. Hence, our theory can be tested by measuring heat current carried by Majorana fermions in the Kitaev magnet coupled to graphene under an electric field.
To this end, let us investigate Majorana heat current given by $j_E = \langle v_E \rangle$ where $v_E=\omega v_x$ is the energy current operator.
Proceeding similarly, we obtain 
\begin{align}
    j_E= \frac{ 5\zeta(3) J^2 e T^3}{4\pi^2  \alpha \delta v^2} E_x.
\end{align}
With the same parameters as used for the estimation of $j$, we obtain $j_E \sim  4 \times 10^{15} \, \text{eV} \, \text{m}^{-1} \cdot \text{s}^{-1}$ which is experimentally detectable.
There may be another contribution to heat current from electrons in graphene. By inserting an insulator between the Kitaev magnet and graphene, this contribution can be reduced.
It is interesting to compare this result with heat current carried by magnons under an applied electric field (magnon-drag thermoelectric effect)\cite{Miura2012}. In this case, the heat current is proportional to $T^{5/2}$ which is distinct from the $T^{3}$ dependence of the  Majorana heat current.
This difference can be attributed to quadratic dispersion relations for magnons and linear dispersion relations for Majorana fermions. 
For magnons in an antiferromagnet with linear dispersions, thermal current carried by the magnons would show the same $T$-dependence as that by Majorana fermions.

The Onsager reciprocal of this effect is the charge current in graphene induced by Majorana current in Kitaev magnets (which can be driven by thermal gradient in the Kitaev magnet by means of the Luttinger potential\cite{Luttinger1964}). This Majorana current-induced charge current can be used as an alternative probe of Majorana current.

When a magnetic field is applied to the Kitaev magnet, the Majorana dispersion can be gapped. The gap is proportional to $h_x h_y h_z/\Delta_s$ where $h_i (i=x, y, z)$ and $\Delta_s$ denote the components of the magnetic field and the spin gap, respectively.\cite{Kitaev2006}
Thus, by applying a magnetic field to the Kitaev magnet, the Majorana current can be strongly suppressed at low temperature due to gap opening. 

In this paper, we assume the absence of visons. Thus, the temperature should be lower than the vison gap $\sim 0.26 J_K$.\cite{Kitaev2006,Knolle2014} For example, for $J_K \sim 10$ meV, the temperature should be lower than 3 meV.

We also assume a weak Kondo coupling ($J$) and performed a perturbative calculation in $J$.
If the Kondo coupling is sufficiently strong, a Kondo effect would occur and the resulting Majorana current would show log $T$ dependence at low temperature. 

In summary, 
we have investigated Majorana current induced by charge current in Kitaev magnet/graphene bilayers via electron-driven Majorana drag effect. 
We have calculated Majorana current density based on the perturbation with respect to a Kondo type Majorana-electron coupling and obtained an analytical expression of the Majorana current induced by charge current, indicative of electronic control of charge-neutral Majorana fermions. We have also discussed Majorana heat current induced by charge current and provided estimations of their magnitudes.

We thank J. Nasu for helpful discussions. 
This work was supported by JSPS KAKENHI Grant No.~JP25K07221.

\section*{Appendix}

1. Derivation of Majorana representation of the Kitaev Hamiltonian.

We introduce the four Majorana fermions $c_i$ and $b_i^\gamma \quad (\gamma = x, y, z)$ at each lattice site $i$ by $s_i^\gamma = i b_i^\gamma c_i$. 
These Majorana operators satisfy $\{c_i, c_j\} = 2\delta_{i,j}, \; \{b_i^\gamma, b_j^\eta\} = 2\delta_{i,j}\delta_{\gamma,\eta} \;$ and $\{c_i, b_j^\eta\} = 0$.
The Fourier transformations of the Majorana conditions give $c_{\boldsymbol{k}}^\dagger = c_{-\boldsymbol{k}}$ and $\{c_{\boldsymbol{k}}, c_{\boldsymbol{k}}^\dagger\} = 1$.
Thus, the operator $c_{\boldsymbol{k}}$ behaves as a complex fermion operator under the condition that $c_{\boldsymbol{k}}$ and $c_{-\boldsymbol{k}}$ are not independent.

Then, the Kitaev Hamiltonian is reduced to 
\begin{eqnarray}
{H}_K = i J_K \sum_{\langle i,j \rangle^\gamma} u_{ij}^\gamma c_i c_j
\end{eqnarray}
with $u_{ij}^\gamma = i b_i^\gamma b_j^\gamma$.

With this Majorana representation, the projection operator and plaquette flux are, respectively,  defined as 
$D_j = -i s_j^x s_j^y s_j^z = b_j^x b_j^y b_j^z c_j$ and
$\phi_p = \prod_{\langle i,j \rangle^\gamma \in p} u_{ij}^\gamma$.
Here, the product is taken over all bonds belonging to the plaquette $p$. 

Because each $u_{ij}^\gamma$ commutes with the Kitaev Hamiltonian ($[\mathcal{H}_K, u_{ij}^\gamma] = 0$), the Hilbert space fragments into sectors labeled by the eigenvalues of $u_{ij}^\gamma$. These eigenvalues are $\pm 1$. This is essentially a static $\mathbb{Z}_2$ gauge field. The fermions ($c_i$) move in the background of these static values.
The transformation from one spin to four Majoranas doubles the local degrees of freedom. The constraint for physical states $D_j |\psi \rangle = |\psi \rangle$ ensures that we only consider states that actually correspond to the original spin system. 
The ground state corresponds to setting $u_{ij}^\gamma = -1$. For the honeycomb lattice, the ground state sector is the vortex-free sector, where the product of the bond variables around any hexagonal plaquette $p$ results in a flux $\phi_p = +1$.

2. Definitions of the Green's functions.

The retarded, advanced, and lesser Majorana Green's functions are defined as
\begin{align}
G_{\alpha\beta}^{R}
(\bm{k};t,t')
&=
-i\theta(t-t')
\left\langle
\left\{
c_{\bm{k}\alpha}(t),
c_{\bm{k}\beta}^{\dagger}(t')
\right\}
\right\rangle ,
\\
G_{\alpha\beta}^{A}
(\bm{k};t,t')
&=
i\theta(t'-t)
\left\langle
\left\{
c_{\bm{k}\alpha}(t),
c_{\bm{k}\beta}^{\dagger}(t')
\right\}
\right\rangle ,
\\
G_{\alpha\beta}^{<}
(\bm{k};t,t')
&=
i
\left\langle
c_{\bm{k}\beta}^{\dagger}(t')
c_{\bm{k}\alpha}(t)
\right\rangle
\end{align}
with $\alpha, \beta=A, B$.
The corresponding electron Green's functions are defined in a similar way.

3. Details of the calculations of Majorana current density.

First, we have 
\begin{equation}
\text{Tr} \, \rho_y G_{\bm{q}}^r(\omega) G_{\bm{q}}^a(\omega) = -\frac{4 v q_x \omega}{[(\omega + i\alpha)^2 - v^2 q^2][(\omega - i\alpha)^2 - v^2 q^2]}.
\end{equation}
Here, the trace is taken over the $\rho$ space.
This is proportional to $q_x$. Thus, the leading term in the Majorana current is proportional to $q_x^2$. Since $E_x = i \Omega {A}_x$, we will expand the self-energy to linear orders in $\Omega$ and $q_x$.

\begin{widetext}
The retarded self-energy $\Sigma^r$ is given by
\begin{align}
    \Sigma^r &= e v_e A_x \sum_{\bm{k}, E} \left[ g_{\bm{k}}(E_-) \tau_x g_{\bm{k}}(E_+) \right]^r g_{\bm{k}-\bm{q}}^<(E') + \left[ g_{\bm{k}}(E_-) \tau_x g_{\bm{k}}(E_+) \right]^< g_{\bm{k}-\bm{q}}^a(E')
\end{align}
with $E'=E-\omega$.
To first orders in $\Omega$ and $q_x$, we have
\begin{align}
    \Sigma^r = -\frac{e v_e^2}{2} \Omega A_x q_x\sum_{\bm{k}, E} f'(E) (g_{\bm{k}}^a(0) - g_{\bm{k}}^r(0)) (g_{\bm{k}}^r(0) \tau_x - \tau_x g_{\bm{k}}^a(0)) g_{\bm{k}}^a(-\omega) \tau_x g_{\bm{k}}^a(-\omega).
\end{align}
Here, we assume $T \ll \delta$ such that the factor $f'(E)$ acts on the electron Green functions as the Dirac's delta function.

Then, we obtain the final expression for $\Sigma^r$:
\begin{equation}
\Sigma^r = \frac{e}{\pi} \Omega  A_x q_x \left[ \frac{\omega(\omega + 2i\delta)(\omega^2 + 2i\delta\omega - 2\delta^2) + 2(\omega + i\delta)^2 \delta^2 \log\left(-\frac{(\omega + i\delta)^2}{\delta^2}\right)}{\omega^3 (\omega + 2i\delta)^3} \right].
\end{equation}

The lesser self-energy $\Sigma^<$ can be calculated as 
\begin{align}
    \Sigma^< &= e v_e A_x \sum_{\bm{k}, E} [ g_{\bm{k}}(E_-) \tau_x g_{\bm{k}}(E_+) ]^< g_{\bm{k}-\bm{q}}^>(E') \\
    &= e v_e  A_x \sum_{\bm{k}, E} \left[ f(E_-) (g_{\bm{k}}^a(E_-) - g_{\bm{k}}^r(E_-)) \tau_x g_{\bm{k}}^a(E_+) + f(E_+) g_{\bm{k}}^r(E_-) \tau_x (g_{\bm{k}}^a(E_+) - g_{\bm{k}}^r(E_+)) \right]  \nonumber \\
    & \times (1 - f(E')) (g_{\bm{k}-\bm{q}}^r(E') - g_{\bm{k}-\bm{q}}^a(E'))\\
= -\frac{e^2 v_e^2}{2} \Omega A_x q_x \sum_{\bm{k}, E} & f'(E) (1 - f(E')) (g_{\bm{k}}^a(0) - g_{\bm{k}}^r(0)) (\tau_x g_{\bm{k}}^a(0) - g_{\bm{k}}^r(0) \tau_x) \nonumber \\
    & \times  \left[ g_{\bm{k}}^a(-\omega) \tau_x g_{\bm{k}}^a(-\omega) - g_{\bm{k}}^r(-\omega) \tau_x g_{\bm{k}}^r(-\omega) \right].
\end{align}

Here, we have expanded $\Sigma^<$ to first orders in $\Omega$ and $q_x$.

Thus, we obtain $\Sigma^<$ as
\begin{equation}
\Sigma^< = 2i \frac{d}{d\omega} (\omega n(\omega)) \, \text{Im} \, \Sigma^r
\end{equation}
with the Bose-Einstein distribution function $n(x) = [\exp(x/T) - 1]^{-1}$.
Here, we have used the relation $\int dE \, f(E + \omega) \frac{df(E)}{dE} = \int dE \, [1 - f(E - \omega)] \frac{df(E)}{dE} = \frac{d}{d\omega} [ \omega n(\omega) ]$.
Hence, we can rewrite the lesser component in Eq. (12) as 
\begin{equation}
    [G_{\bm{q}} \Sigma_{\bm{q}} G_{\bm{q}}]^< = f(\omega) (G^a \Sigma^a G^a - G^r \Sigma^r G^r) + G^r \left( f(\omega) + \frac{d}{d\omega} (\omega n(\omega)) \right) (\Sigma^a - \Sigma^r) G^a
\end{equation}
with the retarded Majorana Green's function in Eq. (3) and the retarded self energy in Eq. (24). Plugging this into Eq. (11), we obtain the expression of the Majorana current density.

\end{widetext}

\end{document}